\documentstyle[preprint,aps]{revtex}
\begin{document}
\tighten
\draft
\preprint{
  \parbox[t]{30mm}{
    hep-th/9802101\\
    DPNU-97-53\\
  }
}
\title{
Dynamical Generation of Fermion Mass and Magnetic Field
 in Three-Dimensional QED with Chern-Simons Term
}
\author{Taichi Itoh \cite{taichi}
  and Hiroshi Kato \cite{kato}}
\address{
  Department of Physics, Nagoya University, Nagoya 464-8602, Japan}
%\date{\today}
\maketitle
\begin{abstract}
We study dynamical symmetry breaking in three-dimensional QED 
with a Chern-Simons (CS) term, 
considering the screening effect of $N$ flavor fermions.
We find a new phase of the vacuum, in which both the fermion mass 
and a magnetic field are dynamically generated, 
when the coefficient of the CS term $\kappa$ equals $N e^2/4 \pi$. 
The resultant vacuum becomes the finite-density state 
half-filled by fermions.
For $\kappa=N e^2/2 \pi$, we find the fermion remains 
massless and only the magnetic field is induced. 
For $\kappa=0$, spontaneous magnetization does not occur 
and should be regarded as an external field.
\end{abstract}


\narrowtext

\indent
Field theoretical models in ($2+1$)-dimensional space-time have 
attracted much attention as effective theories at long 
distance in planar condensed matter physics. 
Especially, quantum electrodynamics in $2+1$ dimensions 
(QED${}_3$) has been intensively studied in connection with the 
effective theories of high-$T_c$ superconductivity \cite{aitch}, 
as well as the probe for ($3+1$)-dimensional quantum chromodynamics. 
In $2+1$ dimensions, there can be a topological gauge action 
known as Chern-Simons (CS) term. 
This term connects a magnetic field $B$ with an electric charge density 
$e\langle\psi^\dagger \psi\rangle$ for fermion field 
$\psi$. From this peculiar property, the CS term is used 
in the field theoretical understanding of 
the fractional quantum Hall effect \cite{frad}. 

As a natural extension of QED${}_3$, a theory 
which gauge field action includes both the CS term and the Maxwell term 
was proposed by Ref.\ \cite{deser}. 
In this theory (CS-QED${}_3$), the coefficient $\kappa$ for CS term 
gives the photon a gauge invariant mass which explicitly violates 
the parity symmetry. 
Some years ago, Hosotani \cite{hosotani} showed 
that spontaneous magnetization occurs in CS-QED${}_3$, 
through breaking the most secret symmetry---Lorentz invariance.
In this theory, the Gauss law 
$\kappa B = -e\langle\psi^\dagger \psi\rangle$ 
follows from the equation of motion. 
Thus the magnetized vacuum corresponds to state 
with finite fermion density $\langle\psi^\dagger \psi\rangle \neq 0$. 
In Ref.\ \cite{itoh} a chemical potential term 
$\mu\, \psi^\dagger \psi$ was introduced as an explicit breaking term 
for Lorentz symmetry, and the condensation 
$\langle\psi^\dagger \psi\rangle$ 
at $\mu \rightarrow 0$ limit was studied. 
It was found that the above vacuum is stable if and only if 
the fermion bare mass is zero.

In this paper, we examine the possibility that Lorentz symmetry is 
broken in a theory in which fermion mass and magnetization are both 
spontaneously generated. 
We regard the fermion bare mass term $-m\,\bar{\psi}\psi$ 
as an explicit breaking term for flavor $U(2N)$ symmetry \cite{appel} 
as well as a chemical potential term 
$\mu\, \psi^\dagger \psi$ for Lorentz symmetry.
Solving the Schwinger-Dyson (SD) equation, 
we clarify whether or not two condensations, 
$\langle\psi^\dagger \psi\rangle$ and $\langle\bar{\psi}\psi\rangle$, 
are dynamically realized at the symmetric limit 
$\mu\rightarrow 0$, $m\rightarrow 0$. 
They signal fermion mass generation and magnetization, respectively. 
For $\kappa=N e^2/4 \pi$, we find a new phase of vacuum, 
in which both massive fermions and a magnetic field are dynamically 
generated so that both symmetries are simultaneously broken. 
The vacuum stability is also examined by calculating 
the Cornwall-Jackiw-Tomboulis (CJT) potential \cite{corn} 
and 2-loop effective potential.

For simplicity, we set Dirac fermion in four components. 
The $\gamma$-matrices are given as follows. 
\begin{equation}
\begin{array}{lll}
\gamma^0\!=\!\left(\!\!
\begin{array}{cc}
\sigma_3 & 0 \\
0 & -\sigma_3 \\
\end{array}\!\!\right),
&\!
\gamma^1\!=\!\left(\!\!
\begin{array}{cc}
i \sigma_1 & 0 \\
0 & -i \sigma_1 \\
\end{array}\!\!\right),
&\!
\gamma^2\!=\!\left(\!\!
\begin{array}{cc}
i \sigma_2 & 0 \\
0 & -i \sigma_2 \\
\end{array}\!\!\right).
\end{array}
\end{equation}
We use the metric ${\rm diag}(g^{\mu \nu})=(-1,1,1)$, 
so that the $\gamma$-matrices satisfy the algebra 
$\{\gamma^{\mu},\gamma^{\nu}\}=-2 g^{\mu \nu}$. 
Our starting Lagrangian including explicit breaking terms is
\begin{eqnarray}
{\cal L} &=&
-\frac{1}{4}{\cal F}_{\mu \nu}{\cal F}^{\mu \nu}
-\frac{\kappa}{2}\epsilon_{\mu \nu \rho}
{\cal A}^{\mu} \partial^{\nu} {\cal A}^{\rho}
-\frac{1}{2 \xi}(\partial {\cal A})^2 \nonumber \\
& & +\bar{\psi}
\left[ \gamma^{\mu}(i\partial_{\mu}+e {\cal A}_{\mu})
-m +\mu \gamma^0 \right]\psi,
\end{eqnarray}
where $\psi$ denotes the $N$ flavor four-component fermion 
and we confine ourselves to the case that fermion mass is 
dynamically generated only as a parity conserving 
mass. From the above Lagrangian, we acquire the gauge field equation 
$\partial_{\nu} {\cal F}^{\nu \mu}
-(\kappa/2)\epsilon^{\mu \nu \rho}{\cal F}_{\nu \rho}
=-e\, \bar{\psi} \gamma^{\mu} \psi$ 
whose vacuum expectation value gives the Gauss law constraint 
$\kappa B = -e \langle\psi^\dagger\psi\rangle$ 
under the existence of constant magnetic field. 

It is known that QED${}_3$ is a super-renormalizable theory and 
its beta function for coupling constant has a nontrivial 
infrared fixed point \cite{appel}. 
Therefore we can not neglect the screening effect of 
vacuum polarization at long distance. 
Following Ref.\ \cite{appel}, we introduce dimensionful coupling 
$\alpha = N e^2/4 \pi$ and keep $\alpha$ finite 
when $N$ is taken to infinity so that 
the vacuum polarization effect can be taken into the effective action 
successively in $1/N$ expansion. 
Since we also attempt to investigate the spontaneous magnetization, 
we separate the gauge field ${\cal A}^\mu$ into back ground 
and propagating fields as 
${\cal A}_\mu = A^{\rm ext}_\mu + A_\mu$, with 
$A^{\rm ext}_\mu (x)= B x_2 \delta_{\mu 1}$. 
The back ground field $A^{\rm ext}_\mu$ can be fully 
contained into the fermion propagator 
\begin{equation}
S(x,y)= -\left\langle x\left|
\frac{1}{\gamma^{\mu}(i\partial_{\mu}+e A^{\rm ext}_{\mu})
-m +\mu \gamma^0}\right|y\right\rangle,
\end{equation} 
by using the proper time method \cite{schwinger}. 
It is determined as
\begin{equation}
S(x,y)=\exp
\left(\frac{ie}{2}(x-y)^{\mu}A_{\mu}^{\rm ext}(x+y)\right)
\widetilde{S}(x-y),\label{fpro1}
\end{equation}
and the Fourier transform for $\widetilde{S}(x-y)$ is given as 
\begin{eqnarray}
\widetilde{S}(k) &=& i\int_{0}^{\infty}\!\!d s\,\exp\left[
-i s \left(m^2 - k_{\epsilon}^2 
+\frac{\tan (eBs)}{eBs}{\bf k}^2
\right)\right] \nonumber \\
&& \times \left\{\left[1+\gamma_{1}\gamma_{2}\tan (eBs)\right]
\left(m + \gamma^{0}k_{\epsilon}\right)\right. 
\nonumber \\ & & \left. 
-(\gamma^{1}k_1 +\gamma^{2}k_2)\sec^{2}(eBs)\right\},\label{fpro2}
\end{eqnarray}
where $k_{\epsilon}:=k^{0}+\mu +i\epsilon\,{\rm sign}(k^{0})$ 
which modifies the $i\epsilon$ prescription to be consistent 
with the shift of Hamiltonian by $\mu$. 
The photon propagator can be read from ${\cal L}$ and is 
\begin{equation}
\Delta^{\mu \nu}(p)=\frac{1}{p^2 + \kappa^2}
\left[g^{\mu \nu}-\frac{p^\mu p^\nu}{p^2}
+i\kappa\frac{p_\rho\epsilon^{\mu\nu\rho}}{p^2}
\right]
+\xi\frac{p^\mu p^\nu}{(p^2)^2}, 
\end{equation}
where we see that the Chern-Simons coefficient 
$\kappa$ plays a role of gauge invariant photon mass.

The effective gauge action improved by fermion 1-loop correction is 
constructed by integrating out the fermion field \cite{jackiw} and 
truncating up to the next to leading order 
in $1/N$ expansion, that is,
\begin{eqnarray}
\Gamma[A] &=& 
-i N\,{\rm Tr Ln}\, S^{-1}
+\int\! d^3 x\left[-\frac{1}{2}B^2 - \kappa B A^{0}(x) \right.
\nonumber\\ && \left.
+\frac{1}{2}A^{\mu}(x)D_{\mu\nu}^{-1}(-i\partial)A^{\nu}(x)
\right],\label{eff}
\end{eqnarray}
where $D^{-1}$ denotes the inverse of the improved photon propagator 
$D_{\mu \nu}(p):= [\Delta^{-1}_{\mu \nu}(p)
-\Pi_{\mu \nu}(p)]^{-1}$.
The vacuum polarization $\Pi_{\mu \nu}(p)$ can be 
regularized in a gauge invariant manner \cite{dittrich,IK}, as 
\begin{eqnarray}
\Pi_{\mu \nu}(p) &=& -\,N e^2 \!\int\!\! \frac{d^3 k}{(2 \pi)^3}\,
{\rm tr}\left[
\gamma_{\mu}\widetilde{S}(k)\gamma_{\nu}\widetilde{S}(k-p)
\right], \nonumber \\
&=& (p_\mu p_\nu -p^2 g_{\mu \nu})\,\Pi_{e}(p)
-i\,p^{\rho}\epsilon_{\mu \nu \rho}\,\Pi_{o}(p) \nonumber \\
& & +\, (p_{\mu}^{\perp} p_{\nu}^{\perp}
-p^2_{\perp} g_{\mu \nu}^{\perp})\,\Pi_{\perp}(p),
\end{eqnarray}
where $p^{\mu}_{\perp}=(0,p_1,p_2)$ and 
${\rm diag}(g_{\mu \nu}^{\perp})=(0,1,1)$.
We notice that the gauge invariant tensor is split into 
the parity conserving part and violating part. 
The latter has the same tensor structure as the CS term, 
or gauge invariant photon mass, in the effective action (\ref{eff}).

Now we will show that the possible values of $\kappa$ under the
existence of the constant magnetic field is restricted by the Gauss law 
$\kappa B = -e \langle\psi^{\dagger}\psi\rangle$. 
It is known that, 
unless $\mu^2 = m^2 + 2 n |eB|$, $n=1,2,\dots\,$, 
the charge condensation 
$\langle\psi^{\dagger}\psi\rangle$
is related to the parity violating part of 
vacuum polarization $\Pi_{o}$ 
with the relation: 
$-e \langle\psi^{\dagger}\psi\rangle = B\, \Pi_{o}(0)$
 \cite{lykken}.
The Gauss law becomes
$B \left[\kappa-\Pi_{o}(0)\right]=0$ which 
means that the nonzero magnetic field can penetrate 
the system when the effective photon mass, 
$\kappa_{\rm eff}:=\kappa-\Pi_{o}(0)$, becomes zero. 
Otherwise, the system with $\kappa \neq \Pi_{o}(0)$ excludes 
a magnetic field whether its origin 
is external or dynamical. 
In this case we should set the magnetic field to zero 
and investigate the dynamical generation of parity breaking 
fermion mass as well as $U(2 N)$ symmetry breaking \cite{kondo}.
We calculate $\Pi_{o}(0)$, when $\mu^2 <m^2 +2|eB|$, as \cite{IK} 
\begin{equation}
\Pi_o (0) = -2\,\alpha\,{\rm sign}(\mu eB)\,\theta (|\mu| -m), 
\end{equation}
where the step function $\theta(x)$ has a value
$1/2$ at $x=0$ as a zero temperature limit of Fermi-Dirac statistics.
We find that $\Pi_o (0)$ has different values according to 
$|\mu|<m$, $|\mu|=m$, and $|\mu|>m$. 
Therefore, the constraint 
$\kappa=\Pi_o (0)$ forces us to take the 
different approach to a symmetric limit $(\mu, m)\rightarrow(0,0)$ 
for each value of $\kappa$ 
so that the nonzero magnetic field can exist. 
The possible way of symmetric limit for each value of $\kappa$  
can be read as 
\begin{equation}
\kappa=\left\{
\begin{array}{ll}
0 & ,|\mu|<m\rightarrow 0 \\ 
-\,\alpha\,{\rm sign}(\mu eB) & ,|\mu|=m\rightarrow 0 \\
-2\,\alpha\,{\rm sign}(\mu eB) & ,m<|\mu|\rightarrow 0\\
\end{array}\right.,\label{kappa}
\end{equation}
which corresponds to empty, half-filled, and fully filled 
lowest Landau levels, respectively. 

In the following, we investigate the dynamical generation of 
fermion mass and magnetic field for theories with
$|\kappa|=0,\alpha,2\alpha$. 
We confirm that the fermion mass $m_d$ and the chemical potential 
$\mu_d$ are dynamically generated at the symmetric limit 
$(m,\mu)\rightarrow(0,0)$. 
In order to be consistent with nonzero magnetic field, 
$m_d$ and $\mu_d$ should satisfy the same relation 
that the explicit breaking parameters $m$ and $\mu$ satisfy 
in Eq. (\ref{kappa}) for each $\kappa$.

It was shown 
in Ref.\ \cite{gusynin} 
that the strong magnetic field played a role of catalyst 
for $U(2N)$ symmetry breaking in the ($2+1$)-dimensional 
Nambu-Jona-Lasinio model. 
Under the strong magnetic field, the wave function for 
a charged fermion is localized around the region with the 
size of magnetic length $l:=1/\sqrt{|eB|}$. 
So, the fermion behaves the same as in a $0+1$ dimension 
and the condensation $\langle\bar{\psi}\psi\rangle$ 
becomes easily formed such as 
in Bardeen-Cooper-Schrieffer theory \cite{gusynin}. 
Recently, Shpagin \cite{shpagin} showed that, by using the SD equation, 
the fermion mass term was dynamically generated for all of 
the number of flavors in QED${}_3$ with an external magnetic field. 
In CS-QED${}_3$, 
when $\kappa$ has the consistent values in Eq. (\ref{kappa}), 
the photon becomes massless and our effective theory, 
described by Eq. (\ref{eff}), is identical to the one in QED${}_3$ 
with an external magnetic field and the Gauss law constraint. 
We, therefore, only have to extend Shpagin's analysis 
using the SD equation so as to include the $\gamma^0$ component 
of fermion self-energy. 

Now we construct the SD equation for fermion self energy.
According to Ref.\ \cite{shpagin}, we assume the strong magnetic field 
$m \ll 1/l$, for which the higher Landau levels 
$\sqrt{m^2 + 2\,n/l^2}$, $n \ge 1$ decouple. 
Thus we only have to treat the lowest Landau level (LLL).
We use the photon propagator improved up to the next to leading order in 
$1/N$ expansion. The SD equation is \cite{shpagin} 
\begin{eqnarray}
G(x,y) &=& S(x,y) - ie^2\int\! d^3 z\,d^3 t\, 
S(x,z) \nonumber\\ 
& & \times\, \gamma^{\mu}G(z,t)\gamma^{\nu}G(t,y)
D_{\mu\nu}(t-z),\label{sd1}
\end{eqnarray}
where we use the bare vertex approximation and 
$G$ denotes the full fermion propagator which should be 
consistently determined through the SD equation. 
We assume that the full propagator $G$ also has the same form 
as $S$ in Eq. (\ref{fpro1}), 
where we only have to replace $S$ with $G$. 
$\widetilde{S}(k)$ in Eq. (\ref{fpro2}) can be decomposed into 
the Landau level poles $k_{\epsilon}^2 = m^2 +2\,n/l^2,\,n=0,1,\dots\,$. 
We see that the higher Landau level poles decouple and 
only the LLL pole, $k_{\epsilon}^2 =m^2$, contributes to 
$\widetilde{S}(k)$ under the strong magnetic field: $m \ll 1/l$. 
Following Ref.\ \cite{shpagin}, we approximate the Fourier transform of
$\widetilde{S}$ and 
$\widetilde{G}$ with its LLL contributions. That is,
\begin{eqnarray}
\widetilde{S}(k) &\simeq& e^{-l^2 k_{\perp}^2}\,
\frac{1}{m-\gamma^0\, k_{\epsilon}}
\left[1-i\gamma^1 \gamma^2 {\rm sign}(eB)\right],\\
\widetilde{G}(k) &\simeq& e^{-l^2 k_{\perp}^2}\,\,\widetilde{g}(k^0)
\left[1-i\gamma^1 \gamma^2 {\rm sign}(eB)\right],
\end{eqnarray}
where $1-i\gamma^1 \gamma^2 {\rm sign}(eB)$ 
is a projection operator to a spin state. 
We notice that the fermion on the lowest Landau level 
essentially behaves like ($0+1$)-dimensional objects. 
 
Under the above approximation, SD equation (\ref{sd1}) is simplified as 
\begin{eqnarray}
\widetilde{g}^{-1}(p^0) &=& m-\gamma^0(p^0+\mu)\nonumber \\ &&
-\frac{ie^2}{(2\pi)^3}\int_{-\infty}^{\infty}\!\! dk^0 \,\gamma^0 \,
\widetilde{g}(k^0)\,\gamma^0 \,\widetilde{D}(p^0 -k^0),\label{sd2}
\end{eqnarray}
where the function $\widetilde{D}$ is defined by
\begin{equation}
\widetilde{D}(p^0):= -\int\! d^2 p_\perp 
\,e^{-l^2 p_{\perp}^{2}/2}\,D_{00}(p^0, p_\perp).
\end{equation}
The function $\widetilde{g}$ is written in the form including the 
scalar component $B$ and $\gamma^0$ component $\widehat{B}$ such as 
\begin{equation}
\widetilde{g}^{-1}(p^0)=B(p^0)-A(p^0)\gamma^0 p^0 
-\gamma^0 [\widehat{B}(p^0)+i \epsilon\,{\rm sign}(p^0)],\label{fng}
\end{equation}
without loss of generality. 
In Eq. (\ref{sd1}), we use the bare vertex approximation, 
so we must set $A(p^0)\equiv 1$ in order to maintain the 
consistency with the Ward-Takahashi identity. 
We also assume $B(p^0)$ is positive definite and 
$B_\pm (p^0):=\widehat{B}(p^0)\pm B(p^0)$ have a definite sign 
irrespective of its argument $p^0$ 
so that we can carry out the Wick rotation $p^0=i\bar{p}$ 
uniquely in the SD equation. 

Setting $A(p^0)\equiv 1$ and substituting Eq. (\ref{fng}) 
into the SD equation (\ref{sd2}), 
we acquire the two independent integral equations 
\begin{equation}
B_{\pm}(\bar{p})=\mu \pm m 
+\frac{e^2}{(2\pi)^3}\int_{-\infty}^{\infty}\!\! d\bar{k}\,
\frac{B_\pm(\bar{k})}
{\bar{k}^{2}+B_{\pm}^{2}(\bar{k})}\,\widetilde{D}(\bar{p}-\bar{k}).
\label{sd3}
\end{equation}
At the limit $\mu\pm m\rightarrow 0$, 
the function $\widetilde{D}$ in Eq. (\ref{sd3}) becomes the one 
calculated from the massless photon propagator and 
has logarithmic behavior in the infrared region of momentum. That is 
$e^2 l\widetilde{D}(\bar{p})\simeq 
-8\pi^2 \alpha_0 \ln |l\bar{p}|$ 
with $\alpha_0 := \alpha l/N(1+c\,\alpha l)$. 
The constant $c$ denotes the parity conserving vacuum polarization effect
$\Pi_e (0)\equiv c\,\alpha l$ and is determined by 
the Riemann zeta function as $c=-6\sqrt{2}\,\zeta(-1/2)\simeq 1.76397$. 

Since the integral in Eq. (\ref{sd3}) is dominated at the infrared region, 
we put $\bar{p} =0$ and can replace $l B_\pm (\bar{k})$ 
by its zero momentum values $\omega_\pm := l B_\pm (0)$. 
The SD equation at $\mu\pm m\rightarrow 0$ finally becomes the gap equations: 
\begin{equation}
\omega_\pm = -\frac{\alpha_0}{\pi}
\int_{-\infty}^{\infty}\!\! ds\,
\frac{\omega_\pm}{s^2 +\omega_\pm^2}\ln |-s|,
\end{equation}
which have the nontrivial solution: 
$|\omega_s|=-\alpha_0 \ln |\omega_s|$,
as well as the trivial one. 
We notice that $\omega_s$ satisfies the condition 
$|\omega_s| \ll 1$, 
which is required to support the LLL approximation, 
since $\alpha_0 < 1$ for any value of $e^2$ and $N$. 
The dynamical variables $m_d$ and $\mu_d$ are given by 
$m_d:= (\omega_{+}-\omega_{-})/2l$, 
$\mu_d:=(\omega_{+}+\omega_{-})/2l$.
According to the values of $\kappa$ in Eq. (\ref{kappa}), 
the consistent solutions are determined as 
\begin{equation}
(m_d,\,\mu_d)=\left\{
\begin{array}{lcl}
(|\omega_s|/l,\,\,0) & \rm{for} & \kappa=0 \\
(|\omega_s|/2l,\,\omega_s /2l) & \rm{for} & |\kappa|=\alpha \\
(0,\,\,\omega_s /l) & \rm{for} & |\kappa|=2\alpha
\end{array}\right..
\end{equation}
The solution for $\kappa=0$ coincides with that of 
Ref.\ \cite{shpagin} on the empty vacuum. 
The solution for $|\kappa|=2\alpha$ corresponds to the one 
in Ref.\ \cite{hosotani} on the fully filled vacuum. 
But, in our case, the massless fermion is shown to be 
dynamically generated. The solution for $|\kappa|=\alpha$ is 
a new one which generates massive fermion as well as
the finite-density vacuum half-filled by fermions.

It was not clear whether or not the solutions of SD equation, 
$m_d$ and $\mu_d$, are energetically favorable 
and the magnetization spontaneously occurs for each $\kappa$. 
So, we have to investigate the vacuum energy. 
We assume the strong coupling $\alpha \gg l^{-1}$ and expand 
the vacuum energy $V(B)$ with respect to $1/\alpha l \sim \sqrt{B/e^{3}}$. 
$V(B)$ is constructed from four parts including the Maxwell energy, 
that is, $V(B)=V_{\rm CJT}(B)+V_F(B)+V_P(B)+B^2/2$. 
$V_{\rm CJT}$ denotes the CJT potential \cite{corn}, 
which gives the energy difference between the nontrivial vacuum 
and the trivial one at the presence of the magnetic field. 
It is evaluated for each $\kappa$, in the LLL approximation, as \cite{IK}
\begin{equation}
V_{\rm CJT}(B)= -\frac{N l}{4\pi}\max\{m_d, |\mu_d|\}|eB|^{3/2}
+{\cal O}(B^2),
\end{equation}
which has a negative value and shows that the nontrivial solutions are 
energetically favorable irrespective of $\kappa$. 

$V_F$ ($V_P$) corresponds to the shift 
of zero-point energy for the fermion (photon) 
induced by the magnetic field at the symmetric limit. 
$V_F$ is deduced from the 1-loop effective action (\ref{eff}) 
as $iN\,{\rm Tr}\ln S^{-1}$ in $\mu\pm m\rightarrow 0$, 
that is \cite{IK}, 
\begin{equation}
 V_F(B) = -\frac{N}{4\pi}|eB|^{3/2} 4\sqrt{2}\,\zeta(-1/2)
  +{\cal O}(B^2).
\end{equation}
$V_P$ is also calculated from the effective action (\ref{eff}) 
as a 2-loop contribution of 
$-(i/2)\,{\rm Tr}\ln D^{-1}$ in $\mu\pm m\rightarrow 0$, 
that is \cite{IK}, 
\begin{equation}
 V_P(B) = -\frac{|\kappa|}{\pi^2}|eB|
  \arctan\left(\frac{2|\kappa|}{\pi\alpha}\right)
   +{\cal O}(B^{3/2}).
\end{equation}
The linear term for $B$ is acquired from $\Pi_o$ in the 
photon propagator, since the energy shift of the photon's vacuum energy 
is an appearance of the difference of effective photon mass in the 
infrared momentum region \cite{hosotani}. 

In $V(B)$, the term proportional to $B^{3/2}$ is dominated by 
that of $V_F(B)$ and has a positive coefficient 
for each $\kappa$ \cite{IK}.
As to the linear term for $B$, it appears for only 
$|\kappa|=\alpha$, $2\alpha$, which corresponds to the charge
condensed vacua, and has a negative coefficient. 
Therefore $V(B)$ has a stationary point at $B\neq 0$ 
for $|\kappa|=\alpha$, $2\alpha$, 
so the magnetic field is dynamically generated. 
For $\kappa=0$, the spontaneous magnetization does not occur and 
we must regard a magnetic field as an external one. 

In summary, we have investigated the dynamical symmetry breaking 
in Chern-Simons QED${}_3$. We have found that both the fermion mass and 
the magnetic field are dynamically generated 
when $|\kappa|=\alpha$ and the corresponding vacuum is 
given as a half-filled lowest Landau level.
This is a new phase of vacuum with nonvanishing fermion mass and broken
Lorentz symmetry.
For $|\kappa|=2\alpha$, we have shown that the fermion remains massless and 
only the magnetic field is induced on the fully filled vacuum.
This is the situation in Ref.\ \cite{hosotani}---vanishing
fermion mass and broken Lorentz symmetry.
This configuration of vacuum has been certified 
to be realized through our argument in this paper. 
It should be noticed that the magnetic field does not necessarily 
enhance the mass generation if the vacuum is fully filled by fermions.
This shows the sharp contrast to the results of
Refs.\ \cite{gusynin,shpagin}, 
which are based on the empty vacuum and correspond to the case $\kappa=0$. 

We would like to thank K. Yamawaki and A. I. Sanda 
for their encouragement and enlightening discussions. 
We also thanks V. P. Gusynin, M. Hashimoto, K. -I. Kondo, 
V. A. Miransky, Y. Nagatani, T. Sato, M. Sugiura, and A. Takamura 
for valuable discussions. 
T. I. is grateful to K. Inoue for his encouragement.

%%%%%%%%%%%%%%%%%%%%%%%%%%%%%%%%%

\end{document}